# Shear induced drainage in foamy yield-stress fluids

J. Goyon[1,2,*], F. Bertrand[1], O. Pitois[2] and G. Ovarlez[1]

[1] *Université Paris Est, Laboratoire Navier, LMSGC, Champs sur Marne, France*

[2] *Université Paris Est, Laboratoire LPMDI, Marne la Vallée, France*

**Abstract:** Shear induced drainage of a foamy yield stress fluid is investigated using MRI techniques. Whereas the yield stress of the interstitial fluid stabilizes the system at rest, a fast drainage is observed when a horizontal shear is imposed. It is shown that the sheared interstitial material behaves as a viscous fluid in the direction of gravity, the effective viscosity of which is controlled by shear in transient foam films between bubbles. Results provided for several bubble sizes are not captured by the $R^2$ scaling classically observed for liquid flow in particulate systems, such as foams and thus constitute a remarkable demonstration of the strong coupling of drainage flow and shear induced interstitial flow. Furthermore, foam films are found to be responsible for the unexpected arrest of drainage, thus trapping irreversibly a significant amount of interstitial liquid.

Foams, which are dispersions of densely packed gas bubbles in liquid, exhibit a large variety of mechanical and dynamical behaviors [1]. It is a rich field of fundamental research, as aqueous foams are often used as model systems for soft matter [2,3]. Moreover, foams are used in lot of industrial applications: gas is mixed in many materials to improve their quality, their performance, or to make them lighter. The materials in which gas is mixed usually exhibit a yield stress to avoid bubbles rising [4-5]. In these foamy systems, homogeneity of the sample can be drastically affected by gravity drainage of the interstitial liquid (and the simultaneous rise of the bubbles), thus justifying the large amount of work devoted to the understanding of drainage laws in foams. Note however that in spite of the significant progress realized in this field, most of the research has been conducted on aqueous foams [6], whereas in most industrial applications the continuous phase of foams are non-newtonian. From a more general point of view, drainage of foams has often been studied separately from other relevant topics, i.e. coarsening and rheology, and only a few studies attempted to elucidate their interdependencies. For example, investigation of the drainage of coarsening foams emphasized the strong and complex coupling existing between these two mechanisms [7,8]. Note that the understanding of this coupling is not yet achieved [9,10,11,12] and this constitutes a limiting stage in the complete modelling of foams' behavior. The coupling of drainage and rheology is also of major importance as foams are sheared at anytime in applications (transport, vibrations). From a fundamental point of view, the interdependency of drainage flow and interstitial flow induced by shear is still an open issue. Although significant progress has been done in the modelling of drainage for static foams, a more general theory, i.e. including the effects of shearing on drainage, is required. Here, we focus on the vertical (gravity) drainage behavior of a model foamy material subjected to a controlled shear in the horizontal plane. The continuous phase is a model complex liquid exhibiting a yield stress. Whereas, due to the continuous phase yield stress, the bubble assembly is jammed at rest, we find that an imposed horizontal shear induces a fast vertical drainage. Then, we show that the coupling of shear and drainage is by far stronger than expected and that the resulting drainage behavior defies the classical results of liquid flow in particulate systems.

We study the stability of wet, monodisperse, foamy, yield-stress fluids of initial bubble volume fraction $\Phi = 71 \pm 1\%$. Three bubble radii are studied: 125, 275 and 500 μm. The interstitial yield stress fluid is an oil in water emulsion, composed of dodecane oil droplets stabilized by surfactants described below. The droplets are $2 \pm 1$ μm in diameter. Nearly monodisperse foams are produced by blowing a mixture of nitrogen with a small amount of perfluorohexan ($C_6F_{14}$) in the surfactant solution through a porous glass frit, the porosity of which fixes the bubble size. $C_6F_{14}$ strongly reduces the coarsening rate [10,13]. Two surfactant solutions are used, both in the emulsion and the foam: the first one is a solution of TetradecylTrimethylAmmonium Bromide (TTAB) at 5g/L, the second one is the same TTAB solution with dodecanol (DOH), at 0.4g/L in the foam and 2g/L in the emulsion which is high enough to saturate the oil droplets. TTAB/DOH interfaces have been shown to be less mobile (more rigid) than pure TTAB ones, thus reducing the drainage flow velocity in aqueous foams [6,14]. In the following, TTAB and TTAB/DOH interfaces will be referred to as 'mobile' and 'rigid' interfaces respectively. The foamy yield stress fluid is obtained by mixing softly the foam and the emulsion, both sharing the same continuous phase to allow their mixing without changing the bubbles and droplets interfacial properties, without any additional air introduction. The incorporation of foam into the emulsion changes the emulsion water content; then, depending on the foam water content, we may add controlled amounts of surfactant solution to the mixture to ensure that in all cases, the oil volume fraction of the interstitial emulsion is 75%, and its yield stress is 13Pa.

In order to study the stability at rest and under shear of the foamy yield stress fluids, we carry out our experiments in a coaxial cylinders Couette rheometer (inner cylinder radius: $r_i$=4.1 cm; outer cylinder radius: $r_o$=5 cm; height of sheared fluid: $H$=11 cm). The use of a wide gap allows studying large bubbles; it results here in moderate stress heterogeneity of 50%. The inner cylinder is rotating, and both cylinders are covered with sandpaper to avoid wall slip. This is the best choice that can be made for the purpose of our study as it imposes shear in the plane perpendicular to drainage (which is thus decoupled from the shear flow). The Couette rheometer is inserted in a Magnetic Resonance Imaging (MRI) setup described in [15,16]. The bubble volume fraction is obtained both in radial and vertical directions from density imaging with an accuracy of 0.2% [17,18]. We study the time evolution of the vertical bubble concentration profiles $\Phi(z)$ for inner cylinder rotational velocity $\Omega$ ranging between 15 and 45rpm, which is high enough to ensure that in all cases, the whole gap is sheared. The local shear rate $\dot{\gamma}(r) = -r\partial(V/r)/\partial r$ at a radius $r$ in the gap can be deduced from the velocity profiles $V(r)$ (Fig. 1b inset). In the following, the results are presented as a function of the macroscopic shear rate $\dot{\gamma}_{macro}$, which is the spatial average of $\dot{\gamma}(r)$. We vary $\dot{\gamma}_{macro}$ between 6 and 18s$^{-1}$.

In Fig. 1a, we first observe that the foamy yield stress fluid is perfectly stable at rest for more than 12 hours. The bubble volume fraction remains constant in time everywhere in the cell. This is consistent with the stability criterion determined for a single bubble embedded in a yield stress fluid [19]: spherical bubbles of 500μm are not expected to move as long as the yield stress of the material is higher than 2.3Pa. This result shows that the yield stress of the interstitial emulsion we use is sufficient to stabilize the foam.

Then we start shearing the material at a given macroscopic shear rate $\dot{\gamma}_{macro}$. Since the bubbles do not move in the material at rest, we have measured MRI density profiles at rest after several flow durations. The evolution of the bubble volume fraction profile of a foamy yield stress fluid under shear is shown in Fig.1a. After introduction in the Couette cell, the sample is homogeneous. When the sample is sheared, bubbles move upwards: $\Phi$ increases up to a maximum $\Phi_{max}$ at the top of the cell and decreases down to 0 at the bottom, the central part remaining at $\Phi_0$, as classically observed in creaming experiments. This bubble rise means that, under shear in the parallel plane, there is no more yield resistance to flow in the vertical direction. We checked that the bubble size and emulsion properties are not changed during the experiments: our observations are the consequence of intrinsic properties of our systems and not artifacts such as bubbles coalescence or oil droplets coalescence. A similar effect was observed for the sedimentation of solid particles in yield stress fluids [20] and granular materials [21].

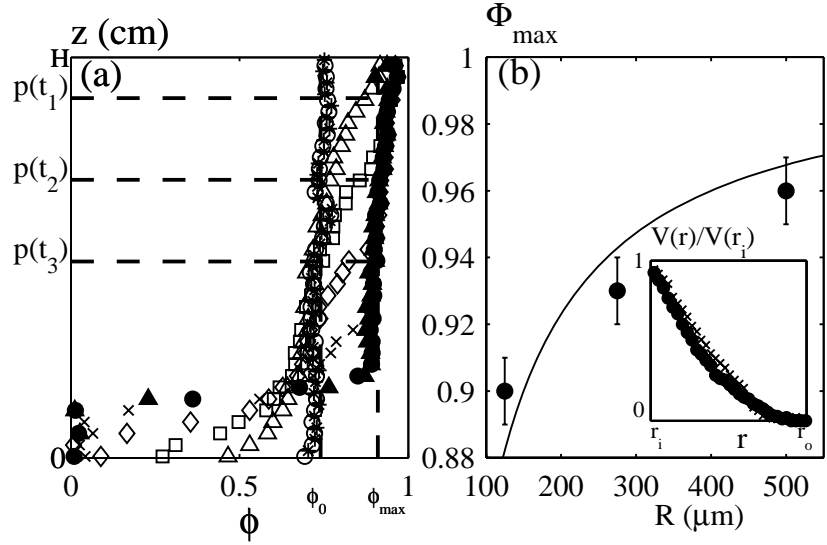

Fig. 1 a) Bubble volume fraction $\Phi$ vs. height $z$ of a foamy yield stress fluid with 275μm radius bubbles, for a 12 hours rest (*), and as function of shear duration at $\dot{\gamma}_{macro}=18s^{-1}$: (O) t=0s, (Δ) 135s, (□) 245s, (◊) 425s, (x) 665s, (●) 965s and (▲) 1265s. (b) $\Phi_{max}$ vs. bubble radius. The solid line is the calculated gas volume fraction assuming that the liquid is contained in films of thickness $h_f=10$ μm (see text). All materials where made with TTAB. Inset: Dimensionless velocity profiles $V(r)/V(r_i)$ as function of the position $r$ in the gap for a sample at $\Phi_0$ (●) and $\Phi_{max}$ (x). The profiles were averaged over 4cm at middle height of the sample; similar results were found in the upper and lower part of the cell.

After some time (of order 1000s in Fig. 1a), a steady state is finally reached: the profiles do not evolve anymore in time. The final sample is composed of pure emulsion at the bottom ($\Phi=0\%$) and of a homogeneous dry foamy emulsion at the top ($\Phi_{max}\approx 93\%$ in Fig.1a). Moreover, we observed that $\Phi_{max}$ depends on the bubble size and surprisingly, the amount of emulsion remaining indefinitely in the foam at this final stage can reach 10% of the total volume (Fig. 1b). Note that a careful inspection of velocity profiles in the dry part (see the inset of Fig. 1b), shows that the whole sample is sheared, so the quantity of emulsion trapped into the foam has to be attributed to an intrinsic retention mechanism.

In order to study in more details the role of the rheological properties of the interstitial yield stress fluid, it is worth studying the kinetics of drainage. From the volume fraction profiles $\Phi(z,t)$, we observe a well defined front for the interstitial emulsion flow; we determine its position as *p(t)* where $\Phi(h,t)=(\Phi_{max}+\Phi_0)/2$ (see Fig. 1a). Note that we checked that the following analysis is not affected by this choice. The bottom of the Couette cell can have disruptive effects, so we analyze the drainage kinetics only in its first stage (when *p* evolves by less than 15% of the sample's total height). In this case, *p(t)* varies linearly in time, as classically observed in creaming experiments, which allows us to define properly a front velocity. By analogy with classical drainage experiments, we focus on the drainage velocity: $V_f = (dp(t)/dt)(\Phi_{max} - \Phi_0)/\Phi_0$. In the following, we study the impact of several control parameters on this velocity: the shear rate, the bubble size and the interface mobility.

The evolution of the front velocity $V_f$ with the macroscopic shear rate $\dot{\gamma}_{macro}$ is shown in Fig. 2. For given bubble size and surfactant solution, we observe that $V_f$ increases basically linearly with $\dot{\gamma}_{macro}$. To explain this feature, we have to understand how the flow of the interstitial fluid in gravity direction depends on shear, which is orthoradial, i.e. to analyse the 3D rheological behavior of the emulsion [20]. Emulsions behave as Herschel-Bulkley fluids [18]; a 3D form of this behavior has been proposed [20,22] $\tau_{ij} = 2d_{ij}(\tau_c + kd^n)/d$, $d = \left(2\sum_{i,j} d_{ij}^2\right)^{1/2}$ is the total shear rate, $d_{ij}$ is the strain rate tensor and $\tau_c$ the yield stress of the material. From this expression, it can be seen that the shearing in any direction affects the shear stresses in all directions. If the shear rate in the interstitial emulsion due to shear in the parallel plane, $\dot{\gamma}_{interstitial} \propto \dot{\gamma}_{macro}$, is small but dominates over the strain rate due to drainage, then the shear resistance $\tau_{rz}$ to a small vertical strain rate $d_{rz}$ induced by drainage reads $\tau_{rz} = 2[\tau_c/\dot{\gamma}_{interstitial}]d_{rz}$, which is that of a purely viscous fluid of effective viscosity $\eta_{eff} \approx \tau_c/\dot{\gamma}_{interstitial} \propto \tau_c/\dot{\gamma}_{macro}$. This analysis is consistent with our observation of a drainage velocity proportional to $\dot{\gamma}_{macro}$. Consistently, it has been recently shown that the sedimentation velocity of a dilute suspension of hard spheres in a sheared yield stress fluid scales with the Stokes velocity in a Newtonian fluid of viscosity $\eta_{eff}$ [20]. Finally, this analysis allows the comparison of drainage velocities with expressions valid for interstitial Newtonian media, the relevant viscosity to consider being $\eta_{eff} \approx \tau_c/\dot{\gamma}_{interstitial}$. The main problem in the following is to properly evaluate $\dot{\gamma}_{interstitial}$ in connection with the foam geometry.

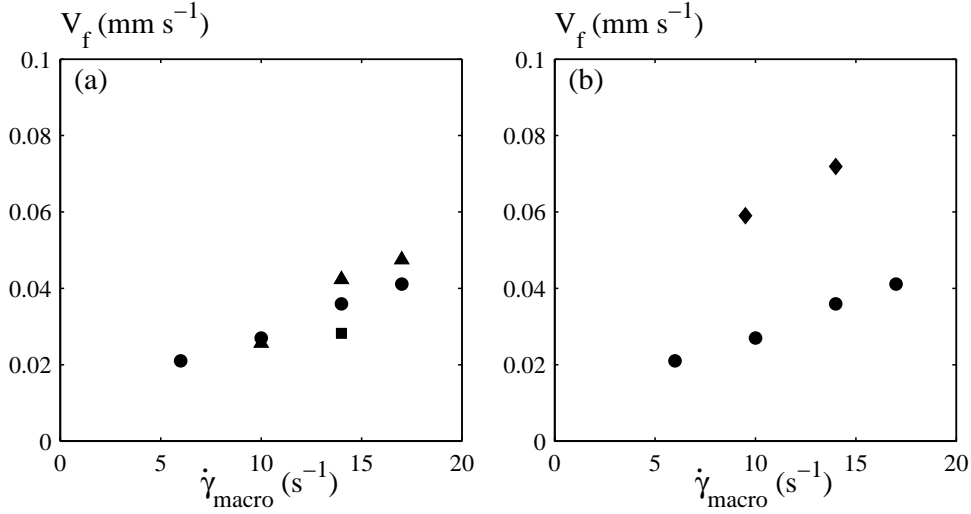

Fig. 2: Drainage front velocities vs. macroscopic shear rate. (a) Influence of the bubbles radius: (■) 125μm, (●) 275μm and (▲) 500μm. (b) Influence of the surface mobility for the 275μm bubbles: (●) TTAB and (♦) TTAB+DOH.

To be more quantitative, we now compare the measured drainage velocity with that measured for liquid flow in unsheared bubbly systems. As discussed above, the interstitial liquid drained through the foam channels behaves as a purely viscous fluid, the viscosity of which depends on the shear rate in the foam channels $\dot{\gamma}_{channels}$. This latter can be estimated from the relative velocity of the bubbles, $2R\dot{\gamma}_{macro}$, and from the diameter of passage of the foam channels, $\approx R/3$ at $\Phi = 0.7$: $\dot{\gamma}_{channels} \approx 6\dot{\gamma}_{macro}$. The effective viscosity of the interstitial liquid in the gravity direction is then conveniently approximated by $\eta_{channels} \approx \tau_c / 6\dot{\gamma}_{macro}$. The velocity for gravity drainage in unsheared bubble systems can be expressed as $V_{th} = \alpha(\Phi) R^2 / \eta_{channels}$ [23], where $\alpha(\Phi)$ is a coefficient depending on both the gas volume fraction and the surface mobility. Recent experiments [23] have provided values for $\alpha$ for both solutions investigated in the present study: $\alpha = 17.8$ and $7$ kg·m$^{-2}$s$^{-2}$ at $\Phi = 0.71$ for mobile and rigid interfaces respectively. For $\dot{\gamma}_{macro} = 14$ s$^{-1}$, the effective viscosity of the interstitial liquid is $\eta_{channels} \approx 0.15$ Pa·s and predicted values of $V_f$ for mobiles interfaces are: $1.8\ 10^{-3}$mm/s for R = 125μm, $9\ 10^{-3}$mm/s for R = 275μm and $3\ 10^{-2}$mm/s for R = 500μm. The predicted drainage velocity compares well with the one measured for the largest bubbles (fig. 2a), but the foam made of smallest bubbles appears to drain much faster than expected, at least by one order of magnitude. Surprisingly, we actually observe in Fig. 2a that the bubble radius has very limited impact on the drainage velocity $V_f$. In aqueous foams, $V_f$ scales as R$^2$ [14] and would change by a factor 16 from the smallest to the largest bubbles; here, it increases by only a factor 1.5.

From now, we propose to interpret these results by considering the flow behavior of the yield stress fluid in foam films. It has been shown that drainage of an isolated horizontal foam film containing colloidal suspensions [24] leads to a decrease of the film thickness down to a critical value below which at least one layer of particles is trapped in the film. In the sheared foamy emulsion, transient films are formed from collisions of neighbouring bubbles [25] and we expect the behavior of those films to be similar to that reported for an isolated film [24]. Thus, it is expected that in the foamy emulsion, foam films always contain a few trapped layers of emulsion droplets that cannot be ejected, thus defining a minimum film thickness $h_f$. The minimum volume fraction associated to those foam films at the end of drainage can be determined using the geometrical model of the well-known Kelvin cell: the films' surface area per bubble is $S_f \approx 14R^2$ [7] and the corresponding film volume is $V_f = S_f h_f / 2 \approx 7 h_f R^2$. The maximum gas volume fraction thus writes $\Phi_{max} \approx 1 - 1/(1 + 0.6 R / h_f)$. As shown in Fig.1b, a very good agreement is found with experimental values assuming $h_f$ = 10μm, which indeed corresponds to a few emulsion droplet diameters. Note that this large value differs by one or two orders of magnitude from the one measured for aqueous foams in draining conditions [9].

We argue that this strikingly high value for the film thickness is also responsible for the surprising drainage behavior reported above. Under shear, during the approach of two neighbouring bubbles, the interstitial liquid is squeezed out from the contact area so that the film thickness decreases at any stage of its existence from its initial value $h_0$ down to the final minimum value $h_f$. During the film thinning process, a significant amount of emulsion is then drained towards connected foam channels, thus increasing the effective flow rate inside those channels. To go one step further, one has to evaluate the emulsion effective viscosity in the films, i.e. to estimate the shear rate $\dot{\gamma}_{films}$ in the films. For dry aqueous foams, shear results in both a squeeze flow and a shear flow in the films [25]; the squeeze flow results in an important dynamic pressure in the films that has to be counterbalanced on the film surface by the bubble capillary pressure, which thus controls (and limits) the squeeze velocity (see Eq. 1 in [25]); it can then be shown that dissipation occurs mainly in the shear flow [25]. Here this is not the case. It should be noted that our systems are initially very wet and that, as shown above, the film are thick ($h$>10μm). Then, the capillary pressure $2\gamma/R$ (from 640 Pa for the 125μm bubbles to 160 Pa for the 500μm, $\gamma$ is the surface tension) has to be compared to the pressure due to the squeeze flow of the interstitial yield stress fluid (the emulsion) at yield [22]: $\frac{2}{3}\tau_c \frac{R_f}{h}$, where $R_f \approx (Rh)^{1/2}$ is the typical size of the film area; this dynamic pressure varies from 30 Pa for the 125μm bubbles to 60 Pa for the 500μm, if $h=h_f=10$μm. This shows that the velocity of the squeeze flow in our systems is not limited by the capillary pressure and thus that the decrease in thickness of a fresh foam film is initially controlled by the relative velocity of the two bubbles, i.e. $-dh/dt \approx 2R\dot{\gamma}_{macro}$. The average liquid velocity at the film/channel transitional area is thus given by: $u \approx (2R\dot{\gamma}_{macro})R_f/\bar{h}$, where $\bar{h}$ an average thickness value during the film thinning, expected to be slightly above $h_f$, i.e. $\bar{h} \approx 15$μm. The resulting squeeze shear rate is $\dot{\gamma}_{films} \approx (2R^2/\bar{h}^2)\dot{\gamma}_{macro}$ (which is dominant over the shear rate of the shear flow in contrast with dry aqueous foams), and the shear rate in the films appears to be much higher than in the channels: $\dot{\gamma}_{films}/\dot{\gamma}_{channels} \geq 20$ for the 125μm bubbles for example. Obviously, the coupling of flows in films and channels is efficient only if the volume of emulsion drained from one film is comparable to that contained in one channel. The film to channel volume ratio is $m \approx 2R_f\bar{h}/\delta R^2 \approx (2/\delta)(\bar{h}/R)^{3/2}$ (where $\delta = \sqrt{3} - \pi/2$ [26]). $m \approx 0.06$ for the 500μm bubbles, so that only a few percent of the liquid in the channels comes from the films: this explains why the measured drainage velocities are in agreement with the first theoretical analysis based on $\dot{\gamma}_{channels}$. In opposition, for the 125μm bubbles, nearly 50% of the liquid contained in the channels comes from the films. As a result, the apparent viscosity of the liquid in the channels is mainly controlled by $\dot{\gamma}_{films}$, and the previous theoretical drainage velocity based on $\dot{\gamma}_{channels}$ is underestimated by a factor $\dot{\gamma}_{films}/\dot{\gamma}_{channels} \approx 20$. This is in good agreement with measured velocities. As a summary, while shearing has no significant effect on the dynamics of aqueous foam drainage [27], it drastically modifies the drainage behavior of the complex liquid: there is an additional contribution of the films on drainage, which increases with decreasing bubble size.

Finally, the effect of surfactant mobility is considered. A comparison of results obtained for 'mobile' and 'rigid' interfaces is presented in Fig. 2b. Surprisingly, the impact of the surfactant on the drainage velocity is opposite to what is usually measured for 'static' aqueous foams [14,28]. Indeed, we observe that drainage is more than 2 times faster for 'rigid' interfaces than for 'mobile' ones, for a given macroscopic shear rate. Qualitatively, this effect can be understood by considering that the shear between bubbles (films and channels) is more efficient for 'rigid' interfaces than for 'mobile' interfaces, leading to a further decrease of the interstitial fluid effective viscosity. Additionally, the surface tension of TTAB/DOH solution is lower than for pure TTAB [10], meaning that bubbles are more easily flattened during collisions and that the discussed above coupling of flows is stronger for TTAB/DOH.

In conclusion, we have studied the stability under gravity of model wet foamy yield stress fluids. We have shown that while the yield stress of the interstitial fluid stabilizes the foam at rest, rapid drainage is induced by shear. The interstitial yield stress fluid then behaves as a viscous fluid in the direction orthogonal to shear, which is that of gravity, and is thus drained. Striking differences with the case of static aqueous foams have been observed, namely the independence of the velocity on the bubble size and the unexpected arrest of drainage at high liquid fraction: these last behaviors can be understood by considering the shear induced flow of interstitial material in the transient foam films and its coupling with the one in the foam channels.

The authors thank Institut Carnot VITRES and the French Space Agency for funding.

* Corresponding author
   Electronic address: julie.goyon@lcpc.fr

[1]  D. Weaire and S. Hurtzler, The physics of foams (Oxford University Press, NY, 2000).

[2]  P. Sollich et al., Phys. Rev. Let., **78**, 2020 (1997).

[3]  R. Höhler and S. Cohen-Addad, J. Phys. Condens. Matter, **17**, R1041 (2005).

[4]  US Patent 6579557.

[5]  S. Guignot et al., Chemical Engineering Science (2010), doi:10.1016/j.ces.2009.12.039.

[6]  A. Saint-Jalmes, Soft Matter **2,** 836 (2006).

[7]  S. Hilgenfeldt et al., Phys. Rev. Lett. **86**, 4704 (2001).

[8]  K. Feitosa and D.J. Durian, Eur. Phys. J. E. **26**, 309 (2008).

[9]  V. Carrier et al., Phys. Rev. E **65**, 061404 (2002).

[10] O. Pitois et al., J. Colloid Interface Sci. **282**, 458 (2005).

[11] O. Pitois et al., Eur. Phys. J. E. **30**, 27 (2009).

[12] N. Louvet et al., J. Colloid Interface Sci. **334**, 82 (2009).

[13] F.G. Gandolfo and H.L. Rosano, J. Colloid Interface Sci. **194**, 31 (1997).

[14] E. Lorenceau et al., Eur. Phys. J. E. **28**, 293 (2009).

[15] J.S. Raynaud et al., J. Rheol. **46** 709 (2002).

[16] S. Rodts et al., C. R. Chim. **7,** 275 (2004).

[17] G. Ovarlez et al., J. Rheol**. 50** 259 (2006).

[18] G. Ovarlez et al., Phys. Rev. E **78** 036307 (2008).

[19] D. Sikorski et al., J. Non-Newtonian Fluid Mech. **102**, 10 (2009)

[20] G. Ovarlez et al., Nature Mat. **9**, 115 (2010).

[21] C. Song et al., PNAS, **102**, 2299 (2005).

[22] P. Coussot. Rheometry of Pastes, Suspensions and Granular Materials (John Wiley & Sons, NY, 2005).

[23] F. Rouyer et al., accepted for publication in Physics of fluids (2010), preprint: http://fr.arxiv.org/abs/0910.2143.

[24] G.N. Sethumadhavan et al., J. Colloid Interface Sci., **240**, 105 (2001).

[25] N. Denkov et al., Phys. Rev. Lett, **100,** 138301 (2008).

[26] S.A. Koehler et al., Langmuir, **16**, 6327 (2000).

[27] R. Soller and S.A. Koehler, Phys. Rev. E **80**, 021504 (2009).

[28] A. Saint-Jalmes et al., Eur. Phys. J. E. **15**, 53 (2004).